\documentclass[conference]{IEEEtran}
\IEEEoverridecommandlockouts
\usepackage{cite}
\usepackage{amsmath,amssymb,amsfonts}
\usepackage{algorithmic}
\usepackage{graphicx}
\usepackage{textcomp}
\usepackage{xcolor}

\usepackage{iftex}
\usepackage{amssymb}
\usepackage{amsmath}
\usepackage{braket}
\usepackage{caption}
\usepackage{subcaption}
\usepackage{comment}
\usepackage{dsfont}
\usepackage{cleveref}
\usepackage{tikz}
\usetikzlibrary{quantikz}
\usepackage{adjustbox}

\newtheorem{definition}{Definition}[section]
\newtheorem{lemma}{Lemma}
\newtheorem{theorem}{Theorem}
\newtheorem{proposition}{Proposition}
\newtheorem{corollary}{Corollary}

\def\BibTeX{{\rm B\kern-.05em{\sc i\kern-.025em b}\kern-.08em
    T\kern-.1667em\lower.7ex\hbox{E}\kern-.125emX}}
\begin{document}

\title{Quantum Distance Calculation for $\varepsilon$-Graph Construction
\thanks{This work is supported by the Agence Nationale de la Recherche projects Q-COAST ANR- 19-CE48-0003, QUACO ANR-17-CE40-0007, and IGNITION ANR-21-CE47-0015, by the project NEASQC that has received funding from the European Union’s Horizon 2020 research and innovation programme under grant agreement No 951821 and by EDF Lab's OSIRIS department which is gratefully acknowledged.}
}

\author{\IEEEauthorblockN{Naomi Mona Chmielewski}
\IEEEauthorblockA{\textit{OSIRIS/L2S} \\
\textit{EDF Lab/CentraleSupélec, Université Paris-Saclay}\\
Palaiseau, France \\
naomi.chmielewski@edf.fr}
\and
\IEEEauthorblockN{Nina Amini}
\IEEEauthorblockA{\textit{L2S} \\
\textit{CNRS, CentraleSupélec, Université Paris-Saclay.}\\
Gif sur Yvette, France \\
nina.amini@centralesupelec.fr}
\and
\IEEEauthorblockN{Paulin Jacquot}
\IEEEauthorblockA{\textit{OSIRIS} \\
\textit{EDF Lab}\\
Palaiseau, France \\
paulin.jacquot@edf.fr}
\and
\IEEEauthorblockN{Joseph Mikael}
\IEEEauthorblockA{\textit{OSIRIS} \\
\textit{EDF Lab}\\
Palaiseau, France \\
joseph.mikael@edf.fr}
}

\maketitle

\begin{abstract}
In machine learning and particularly in topological data analysis, $\varepsilon$-graphs are important tools but are generally hard to compute as the distance calculation between $n$ points takes time $\mathcal{O}(n^2)$ classically. Recently, quantum approaches for calculating distances between $n$ quantum states have been proposed, taking advantage of quantum superposition and entanglement. We investigate the potential for quantum advantage in the case of quantum distance calculation for computing $\varepsilon$-graphs. We show that, relying on existing quantum multi-state SWAP test based algorithms, the query complexity for correctly identifying (with a given probability) that two points are not $\varepsilon$-neighbours is at least $\mathcal{O}(n^3/ \ln{n})$, showing that this approach, if used directly for $\varepsilon$-graph construction, does not bring a computational advantage when compared to a classical approach.
\end{abstract}


\section{Introduction}
Nearest-neighbour or similarity graphs are an important tool in machine learning. They are used in collaborative filtering for recommendation systems \cite{recommenderSystem}, clustering \cite{clustering} and pattern recognition \cite{patternRecognition}. $\varepsilon$-nearest neighbour graphs in particular can be used in the construction of the Vietoris-Rips complex, an important step in Topological Data Analysis (TDA) \cite{ZomorodianVRconstruction}, which has been shown to have potential quantum advantages in certain cases \cite{qtda}, \cite{lgz}.

In order to construct an $\varepsilon$-graph from a point cloud with $n$ points, we need a way to calculate euclidean distances between points. Classically, the distance calculation for the construction of an $\varepsilon$-graph can be done in time $\mathcal{O}(n\log n)$ on average, using the \textit{kd-tree} algorithm \cite{kdtree} which only scans relevant areas in the embedding space. Constructing the full distance matrix between all $n$ inputs takes time $\mathcal{O}(n^2)$.  

Quantumly, distance calculation between two quantum states is commonly done using the \textit{SWAP}-test \cite{swap}, requiring one CSWAP gate and one ancillary qubit. In \cite{multiswap}, the authors propose a method to calculate the distances between all $n$ input states using only $\mathcal{O}(n)$ CSWAP gates and $\mathcal{O}(\log n)$ ancillaries, suggesting an interesting alternative to the classical methods. 

The authors in \cite{multiswapliu} propose a modification of the multi-state SWAP test from \cite{multiswap}, potentially reducing the number of circuit repetitions by combining multiple SWAP tests in parallel, while increasing the number of CSWAP gates in the circuit to $\mathcal{O}(n\log n)$.

In \cite{qnearestneighbor}, the authors suggest a quantum algorithm for finding the nearest neighbour of a query input state by combining amplitude estimation and the Dürr-Hoyer minimisation algorithm \cite{durrhoyer}. The algorithm in \cite{qnearestneighbor} can be used to find the $k$ nearest neighbours, $k\in\mathbb{N}$ by repeating the process $k$ times while removing all previously found nearest neighbours from the data. This approach does not easily extend to $\varepsilon$-nearest neighbour identification.

In this paper, we show that the number of oracle calls necessary to construct an $\varepsilon$-graph using the quantum algorithm proposed in \cite{multiswap} to $(1-\gamma)$ correctly identify that two points are not $\varepsilon$-neighbours is at least $\mathcal{O}(n^3/\ln n)$, where $n$ is the number of input states. 

This paper is organised as follows: in section \ref{sec:quantumdistance}, we recall the definition of an $\varepsilon$-graph for our purposes as well as the standard SWAP test and summarise the algorithm developed in \cite{multiswap}. We also propose a simple extension of the algorithm to $d$-dimensional inputs. We give our main results for the query and gate complexity of the algorithm proposed in \cite{multiswap} in section \ref{sec:oraclecomplexity} by establishing sharp lower bounds for the number of oracle calls necessary to create $\varepsilon$-graphs. Section \ref{sec:discussion} provides a discussion on the presented results. 

\section{Distance Calculation in Quantum Parallel}
\label{sec:quantumdistance}

\subsection{Classical Distance Calculation}
\label{sec:ckassicaldistance}

\begin{definition}[$\varepsilon$-graph]
    Given a finite set of $d$-dimensional points $S \subseteq \mathbb{R}^d$ of size $|S|=n$ and scale $\varepsilon >0$,
     the $\varepsilon$-graph is an undirected graph $G_{\varepsilon}=(V, E_{\varepsilon})$ where $V=S$ and
    \begin{equation*}
        E_{\varepsilon} = \{\{u,v\} \, | \, \delta(u,v)<\varepsilon, u\neq v\in S\}
    \end{equation*}
    where $\delta$ is the euclidean metric.
\end{definition}

Thus, to construct an $\varepsilon$-graph from a given set $S$, for each $u\in S$ we must find all points $v_j \in S$ that are at a distance $d(u, v_j)<\varepsilon$. One way to do this is to use a \textit{brute-force} algorithm that calculates all $n(n-1)/2$ distances between the $n$ points and selects pairs that are within $\varepsilon$-distance. This algorithm takes $\mathcal{O}(n^2)$ time and is linear in the dimensionality $d$.

For large $n$, there are more efficient algorithms such as the \textit{kd-tree} which takes time $\mathcal{O}(n\log n)$ on average by sorting the data into a tree and searching only subsections of that tree \cite{kdtree}. This algorithm thus does not generally calculate all $n(n-1)/2$ possible distances. Note that the dimensionality dependence of the kd-tree algorithm is at least $\mathcal{O}(2^d)$ due to the curse of dimensionality, making the algorithm less suitable for high-dimensional data \cite{arya}.

As the number $n$ of input points tends to be extremely high, it would be favourable to reduce the time dependence in $n$ while also reducing the dependence in $d$. A naive approach to try and reduce the complexity could be to use quantum algorithms that estimate distances in parallel. As the number of CSWAP gates in the algorithm proposed in \cite{multiswap} is $\mathcal{O}(n)$, and a simple extension to $d$-dimensional input states only leads to a linear increase in the number of CSWAP gates, this might be achieved if the query complexity is sufficiently low. In our analysis, we find that the query complexity is too high to achieve any improvement over classical exact algorithms.

\subsection{SWAP-Test}
\label{sec:swaptest}

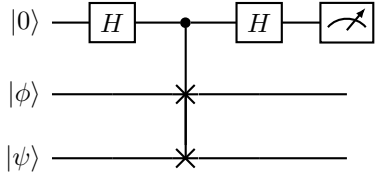
\begin{figure}[!t]
    \centering
    \begin{quantikz}
    \lstick{$\ket{0}$} & \gate{H} & \ctrl{2} & \gate{H} & \meter{} \\
    \lstick{$\ket{\phi}$} & \qw & \targX{} & \qw & \qw \\
    \lstick{$\ket{\psi}$} & \qw & \swap{-1} & \qw & \qw
    \end{quantikz}
    \caption{Quantum circuit representing the standard SWAP test}
    \label{fig:swap}
\end{figure}

The common way to calculate the distance between two quantum states $\ket{\phi}$, $\ket{\psi}$ of dimension $d=1$ is done via the SWAP test \cite{swap} (see \Cref{fig:swap}).

The resulting state is then 
    \begin{equation*}
        \frac{1}{2}\ket{0}(\ket{\phi,\psi} + \ket{\psi,\phi}) + \frac{1}{2}\ket{1}(\ket{\phi,\psi} - \ket{\psi,\phi})
    \end{equation*}
so that at the end, the probability of measuring $0$ is 
    \begin{equation*}
    \begin{aligned}
        &\frac{1}{2}(\bra{\phi}\bra{\psi} + \bra{\psi}\bra{\phi})\frac{1}{2}(\ket{\phi}\ket{\psi} + \ket{\psi}\ket{\phi}) \\
        &= \frac{1}{2} + \frac{1}{2}|\braket{\psi|\phi}|^2 \ .
    \end{aligned}
    \end{equation*}
Thus, $\mathbb{P}(0) = 1$ if and only if $\ket{\psi}$ and $\ket{\phi}$ are parallel, and $\mathbb{P}(0) = 1/2$ if and only if $\ket{\psi}$ and $\ket{\phi}$ are orthogonal. Note in particular that $\mathbb{P}(0)\equiv p\in[1/2,1]$. The way to estimate the distance is then to repeat the circuit $N$ times and to take $\hat{p}=1-\frac{1}{N}\sum_{i=1}^N X_i$ as the estimate for the probability, where $X_i$ is a Bernoulli random variable with parameter $q=1-p$, representing the \textit{i}th measurement outcome of the circuit. 
Finally, the distance is estimated by noting that for normalised vectors, we have 
    \begin{equation*}
        |\phi-\psi| = \sqrt{2(1-|\braket{\phi|\psi}|)} = \sqrt{2(1-\sqrt{2p-1})}\ .
    \end{equation*}
In the case where $d>1$, write $\ket{\phi} = \ket{\phi_1}\ket{\phi_2}\cdots\ket{\phi_d}$ and $\ket{\psi} = \ket{\psi_1}\ket{\psi_2}\cdots\ket{\psi_d}$. Then, one CSWAP gate is applied to each pair $\ket{\phi_i}, \ket{\psi_i}$ while the rest of circuit remains the same. Unless otherwise specified, we assume $d=1$ for the remainder of the paper, as the explicit dependence in $d$ is linear and all results are easily extended to $d>1$. %

\subsection{Naive Extension to $n$ States}
\label{sec:naivemultiswap}

A simple way to extend the SWAP test to $n$ input states is to create one SWAP test circuit for each of the $n(n-1)/2$ possible distances. This circuit requires $n(n-1)/2$ CSWAP gates. Since the ancilla qubit is measured at the end of each circuit run, we can reuse the same qubit for the ancillary state. We show in section \ref{sec:oraclecomplexity} that the number of necessary circuit repetitions to $(1-\gamma)$ correctly identify that two points $\ket{\phi_i}, \ket{\phi_j}, i,j\in\{1,\ldots,n\}$ are not $\varepsilon$-neighbours is $\mathcal{O}(n^2)$ for this circuit design, meaning that the total number of CSWAP gates is $\mathcal{O}(n^4)$ for any $\gamma\in(0,1)$.

\subsection{Multi-State SWAP Test}
\label{sec:multiswap}

The authors in \cite{multiswap} propose a recursive gate arrangement to calculate all distances between $n$ quantum states using $\mathcal{O}(n)$ CSWAP gates and $\mathcal{O}(\log n)$ ancillary qubits.

The algorithm is based on a unitary $\mathcal{U}_4$ constructed from $3$ CSWAP gates and $3$ ancillary qubits and is applied to $4$ input states (see \Cref{fig:U4large}). The circuit puts all possible pairs between the $4$ input states into the first two input registers in superposition.

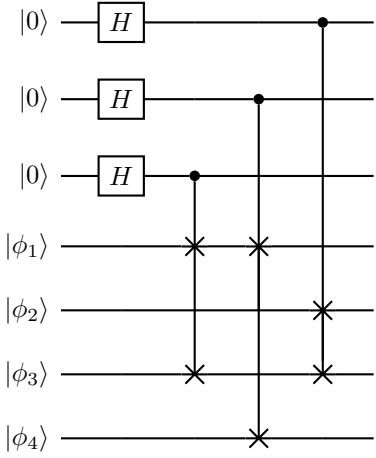
\begin{figure}[!t]
    \centering
    \begin{minipage}{.5\textwidth}
        \centering
        \begin{quantikz}
        \lstick{$\ket{0}$} & \gate{H} & \qw & \qw & \ctrl{5} & \qw \\
        \lstick{$\ket{0}$} & \gate{H} & \qw & \ctrl{3} & \qw & \qw \\
        \lstick{$\ket{0}$} & \gate{H} & \ctrl{2} & \qw & \qw & \qw \\
        \lstick{$\ket{\phi_1}$} & \qw & \targX{} & \targX{} & \qw & \qw \\
        \lstick{$\ket{\phi_2}$} & \qw & \qw & \qw & \targX{} & \qw \\
        \lstick{$\ket{\phi_3}$} & \qw & \swap{-1} & \qw & \swap{-1} & \qw \\
        \lstick{$\ket{\phi_4}$} & \qw & \qw & \swap{-3} & \qw & \qw
        \end{quantikz}
        \caption{Quantum circuit representing $\mathcal{U}_4$ (from \cite{multiswap})}
        \label{fig:U4large}
    \end{minipage}%
\end{figure}
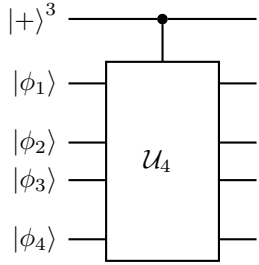
\begin{figure}[!htb]
    \begin{minipage}{0.5\textwidth}
        \centering
        \begin{quantikz}
        \lstick{$\ket{+}^3$} & \ctrl{1} & \qw \\
        \lstick{$\ket{\phi_1}$} & \gate[4][1.5cm]{\begin{minipage}{0.5cm}$\mathcal{U}_4$\end{minipage}} &\qw \\
        \lstick{$\ket{\phi_2}$} & & \qw \\
        \lstick{$\ket{\phi_3}$} & & \qw \\
        \lstick{$\ket{\phi_4}$} & & \qw   
        \end{quantikz}
        \caption{Compact representation of $\mathcal{U}_4$}
        \label{fig:U4small}
    \end{minipage}
\end{figure}
For $n=2^k$ input states, the circuit $\mathcal{U}_n$ that moves all possible pairs into the first two input registers is constructed recursively from the circuit $\mathcal{U}_4$, using $3n/2-3$ CSWAP gates and $d_n = 3\log_2(n/2)$ ancillary qubits (see \Cref{fig:Unlarge}).
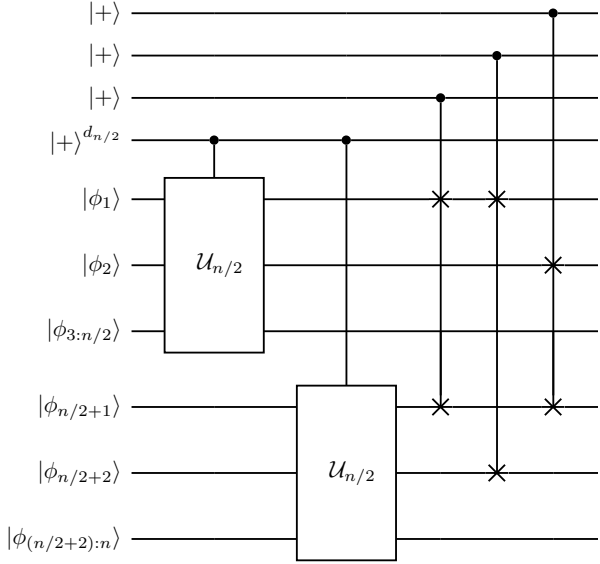
\begin{figure}[!t]
    \begin{adjustbox}{width=0.9\columnwidth,center}
    \centering
    \begin{minipage}{.5\textwidth}
        \centering
        \begin{quantikz}
        \lstick{$\ket{+}$} & \qw & \qw & \qw & \qw & \ctrl{7} & \qw \\
        \lstick{$\ket{+}$} & \qw & \qw & \qw & \ctrl{6} & \qw & \qw \\
        \lstick{$\ket{+}$} & \qw & \qw & \ctrl{5} & \qw & \qw & \qw \\
        \lstick{$\ket{+}^{d_{n/2}}$} & \ctrl{1} & \ctrl{5} & \qw & \qw & \qw & \qw \\
        \lstick{$\ket{\phi_1}$} & \gate[3][1.5cm]{\begin{minipage}{0.5cm}
            $\mathcal{U}_{n/2}$
        \end{minipage}} & \qw & \targX{} & \targX{} & \qw & \qw \\
        \lstick{$\ket{\phi_2}$} &  & \qw & \qw & \qw & \targX{} & \qw \\
        \lstick{$\ket{\phi_{3:n/2}}$} &  & \qw & \qw & \qw & \qw & \qw \\
        \lstick{$\ket{\phi_{n/2+1}}$} & \qw & \gate[3][1.5cm]{\begin{minipage}{0.5cm}
            $\mathcal{U}_{n/2}$
        \end{minipage}} & \swap{-1} & \qw & \swap{-1} & \qw \\
        \lstick{$\ket{\phi_{n/2+2}}$} & \qw &  & \qw & \swap{-1} & \qw & \qw \\
        \lstick{$\ket{\phi_{(n/2+2):n}}$} & \qw &  & \qw & \qw & \qw & \qw
        \end{quantikz}
        \caption{Quantum circuit representing $\mathcal{U}_n$ (from \cite{multiswap})}
        \label{fig:Unlarge}
    \end{minipage}%
    \end{adjustbox}
\end{figure}
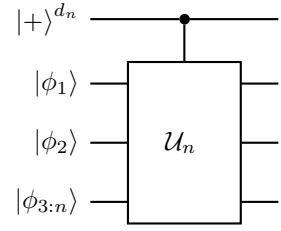
\begin{figure}[!htb]
    \begin{minipage}{0.5\textwidth}
        \centering
        \begin{quantikz}
        \lstick{$\ket{+}^{d_n}$} & \ctrl{1} & \qw \\
        \lstick{$\ket{\phi_1}$} & \gate[3][1.5cm]{\begin{minipage}{0.5cm}
            $\mathcal{U}_n$
        \end{minipage}} &\qw \\
        \lstick{$\ket{\phi_2}$} & & \qw \\
        \lstick{$\ket{\phi_{3:n}}$} & & \qw \\
        \end{quantikz}
        \caption{Compact representation of $\mathcal{U}_n$}
        \label{fig:Unsmall}
    \end{minipage}
\end{figure}

This circuit can be used for any $n$ inputs by simply padding the remaining input register with $\ket{0}$s.

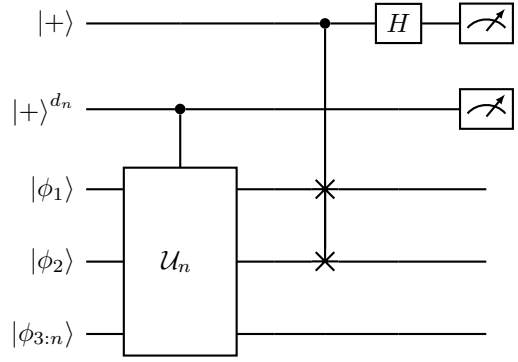
\begin{figure}[!t]
    \centering
    \begin{minipage}{0.5\textwidth}
        \centering
        \begin{quantikz}
        \lstick{$\ket{+}$} & \qw & \qw & \ctrl{3} & \gate{H} & \meter{}\\
        \lstick{$\ket{+}^{d_n}$} & \ctrl{1} & \qw & \qw & \qw & \meter{}\\
        \lstick{$\ket{\phi_1}$} & \gate[3][1.5cm]{\begin{minipage}{0.5cm}
            $\mathcal{U}_n$
        \end{minipage}} &\qw & \targX{} & \qw & \qw\\
        \lstick{$\ket{\phi_2}$} & & \qw & \swap{-1} & \qw & \qw\\
        \lstick{$\ket{\phi_{3:n}}$} & & \qw & \qw & \qw & \qw\\
        \end{quantikz}
        \caption{Final SWAP test and measurement (from \cite{multiswap})}
        \label{fig:fullMultiswap}
    \end{minipage}
\end{figure}

To finish the multi-state SWAP test, one additional ancillary qubit is added, and a final CSWAP gate and Hadamard gate are implemented, analogous to the two-state SWAP test. As in the two-state SWAP test, the top ancilla qubit is measured, and to extract information on all state-pairs, an additional measurement is done on the $d_n$ ancilla qubits (see \Cref{fig:fullMultiswap}).

The quantity to be estimated is then $p_{0ij}$ for $i,j \in \{1, \ldots, n\}$ where $p_{0ij}$ designates the probability of measuring the ancillary state $\ket{0ij}$, where $\ket{ij}$ is shorthand for the $d_n$-dimensional ancillary basis state associated with the couple $(\ket{\phi_i}, \ket{\phi_j})$. As the authors state in \cite[Theorem 2.2]{multiswap}, the probability can be expressed as

        \begin{equation*}
            p_{0ij} = \frac{1+|\braket{\phi_i|\phi_j}|^2}{2^{d_n}}
        \end{equation*}
        where $d_n$ is the number of ancillary qubits $d_n = 3\log(n/2)$, thus 
        
        \begin{equation}
        \label{eq:p0ij}
            p_{0ij} = \frac{2^3(1+|\braket{\phi_i|\phi_j}|^2)}{n^3}\ .
        \end{equation}

\subsection{Extension to Multi-Dimensional Inputs}
\label{sec:higherdim}
        
\begin{figure}[!t]
    \begin{adjustbox}{width=0.9\columnwidth,center}
        \begin{quantikz}
        \lstick{$\ket{+}$} & \qw & \qw & \ \ldots\ \qw & \qw & \qw & \qw & \ \ldots\ \qw & \qw & \ctrl{7} & \ctrl{8} & \ \ldots\ \qw & \ctrl{11} & \qw \\
        \lstick{$\ket{+}$} & \qw & \qw & \ \ldots\ \qw & \qw & \ctrl{3} & \ctrl{3} & \ \ldots\ \qw & \ctrl{5} & \qw & \qw & \ \ldots\ \qw & \qw & \qw \\
        \lstick{$\ket{+}$} & \ctrl{2} & \ctrl{2} & \ \ldots\ \qw & \ctrl{2} & \qw & \qw & \ \ldots\ \qw & \qw & \qw & \qw & \ \ldots\ \qw & \qw & \qw \\
        \lstick{$\ket{\phi_1^1}$} & \targX{} & \qw & \ \ldots\ \qw & \qw & \targX{} & \qw & \ \ldots\ \qw & \qw & \qw & \ \ldots\ \qw & \qw & \qw & \qw\\
        \lstick{$\ket{\phi_1^2}$} & \qw & \targX{} & \ \ldots\ \qw & \qw & \qw & \targX{} & \ \ldots\ \qw & \qw & \qw & \qw & \ \ldots\ \qw & \qw & \qw \\
        \vdots \\
        \lstick{$\ket{\phi_1^d}$} & \qw & \qw & \ \ldots\ \qw & \targX{} & \qw & \qw & \ \ldots\ \qw & \targX{} & \qw & \qw & \ \ldots\ \qw & \qw & \qw\\
        \lstick{$\ket{\phi_2^1}$} & \qw & \qw & \ \ldots\ \qw & \qw & \qw & \qw & \ \ldots\ \qw & \qw & \targX{} & \qw & \ \ldots\ \qw & \qw & \qw \\
        \lstick{$\ket{\phi_2^2}$} & \qw & \qw & \ \ldots\ \qw & \qw & \qw & \qw & \ \ldots\ \qw & \qw & \qw & \targX{} & \ \ldots\ \qw & \qw & \qw \\
        \vdots \\
        \lstick{$\ket{\phi_2^d}$} & \qw & \qw & \ \ldots\ \qw & \qw & \qw & \qw & \ \ldots\ \qw & \qw & \qw & \qw & \ \ldots\ \qw & \targX{} & \qw \\
        \lstick{$\ket{\phi_3^1}$} & \swap{-7} & \qw & \ \ldots\ \qw & \qw & \qw & \qw & \ \ldots\ \qw & \qw & \swap{-5} & \qw & \ \ldots\ \qw & \qw & \qw \\
        \lstick{$\ket{\phi_3^2}$} & \qw & \swap{-10} & \ \ldots\ \qw & \qw & \qw & \qw & \ \ldots\ \qw & \qw & \qw & \swap{-5} & \ \ldots\ \qw & \qw & \qw \\
        \vdots \\
        \lstick{$\ket{\phi_3^d}$} & \qw & \qw & \ \ldots\ \qw & \swap{-11} & \qw & \qw & \ \ldots\ \qw & \qw & \qw & \qw & \ \ldots\ \qw & \swap{-3} & \qw \\
        \lstick{$\ket{\phi_4^1}$} & \qw & \qw & \ \ldots\ \qw & \qw & \swap{-12} & \qw & \ \ldots\ \qw & \qw & \qw & \qw & \ \ldots\ \qw &  \qw & \qw \\
        \lstick{$\ket{\phi_4^2}$} & \qw & \qw & \ \ldots\ \qw & \qw & \qw & \swap{-13} & \ \ldots\ \qw & \qw & \qw & \qw & \ \ldots\ \qw & \qw & \qw \\
        \vdots \\
        \lstick{$\ket{\phi_4^d}$} & \qw & \qw & \ \ldots\ \qw & \qw & \qw & \qw & \ \ldots\ \qw & \swap{-14} & \qw & \qw & \ \ldots\ \qw & \qw & \qw \\
        \end{quantikz}

    \end{adjustbox}
         \caption{Quantum circuit representing $\mathcal{U}_4$ for $d>1$}
        \label{fig:U4d}
\end{figure}
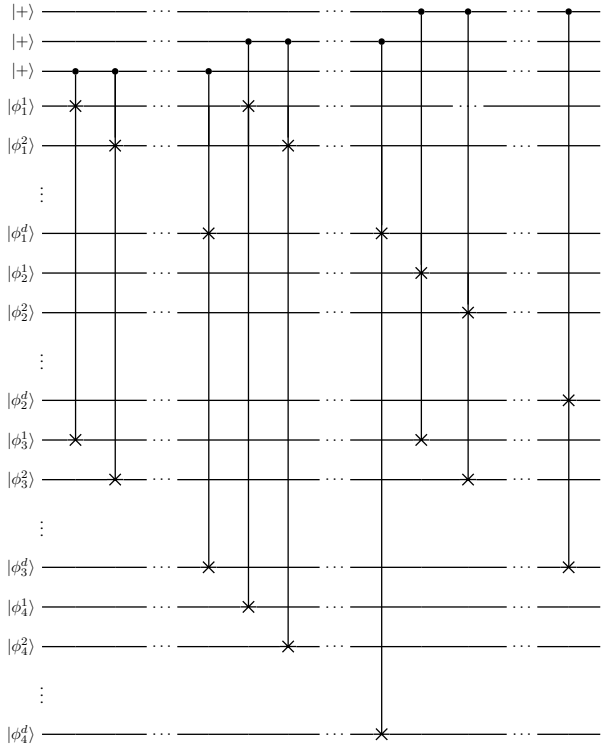

Although the authors only detail their algorithm for one-dimensional inputs, a natural extension to higher dimensional inputs $d>1$ can be done analogous to the simple SWAP test, adding one additional CSWAP gate per dimension for each CSWAP gate present in the base circuit $\mathcal{U}_4$ as well as one additional CSWAP gate per dimension in the final part before the measurement (see Fig \ref{fig:U4d}). This would lead to a linear increase in CSWAP gates, more specifically, the number of gates for $n$ inputs of dimension $d$ is $(3n/2-3)d$, and the number of ancillary states does not change. 

\section{Query Complexity of the Multi-State SWAP Test for $\varepsilon$-Graph Construction}
\label{sec:oraclecomplexity}

Denote by $\mathcal{O}_n$ an oracle that generates $n$ quantum states $\ket{\phi_1},\ldots,\ket{\phi_n}$, for example a qRAM \cite{qram}. In \cite{multiswap}, the authors establish an upper bound on the number of oracle calls necessary to obtain $\gamma$-close estimates of all distances, where $\gamma$-close refers to the expectation of the difference between the estimates and the true distances being at most $\gamma$. The theorem is stated for a larger class of algorithms that includes $\mathcal{U}_n$. We will only focus on $\mathcal{U}_n$ in this paper. Adapting their notations to fit ours, we note $\mu = \{|\braket{\phi_i|\phi_j}|^2\}_{i<j=1}^n$ the ground truth overlaps between all states (instead of $\delta$), $\hat{\mu}=\{\mu_{ij}\}_{i<j=1}^n$ the estimate of the overlaps (instead of $\hat{\delta}$), $n$ the number of input states (instead of $m$), $\gamma\in(0,1)$ the precision parameter (instead of $\epsilon$), and noting that $2^{2d_n}=n^6/2^6$, \cite[Theorem 2.3]{multiswap} states the following:

\bigskip

\begin{theorem}[\cite{multiswap}]
    The circuit $\mathcal{U}_n$ needs at most $\mathcal{O}(n^6/\gamma^2)$ calls to the oracle $\mathcal{O}_n$ to obtain an estimate $\hat{\mu}$ of $\mu = \{|\braket{\phi_i|\phi_j}|^2\}_{i<j=1}^n$ such that $\mathbb{E}[||\hat{\mu}-\mu||_2]\leq \gamma$, where $||\hat{\mu}-\mu||_2^2 \equiv \sum_{i<j}(\hat{\mu}_{ij}-\mu_{ij})^2$.
\end{theorem}

\bigskip

Note in particular that this implies a number $\mathcal{O}(n^7/\gamma^2)$ of CSWAP gates.

To calculate the $\varepsilon$-graph, it is not necessary to calculate all distances exactly. In fact, it suffices to decide whether the distance between any two states is likely smaller than $\varepsilon$.

Define 
\begin{equation*}
    \delta \equiv \mathds{1}_{|\phi-\psi|<\varepsilon}=\mathds{1}_{p> \frac{(1-\varepsilon^2/2)^2+1}{2}}
\end{equation*}
 where $\varepsilon\in\mathbb{R}_+$. For ease of notation, write 
 \begin{equation*}
     \alpha_{\varepsilon} = \frac{(1-\varepsilon^2/2)^2+1}{2} \ .
 \end{equation*}
 Then, $X$ is the binomial random variable $\sim$ $\text{Bin}(N, q)$ defined by $X=\sum_{i=1}^N X_i$ where $X_i$ is a Bernoulli random variable $\sim \text{Ber}(q), q=\mathbb{P}(X_i=1)=1-p$, $p\in[1/2,1)$. Denote by $\hat{p}=1-\sum_{i=1}^NX_i/N$ the estimator of $p$. A \textit{false negative} is identified when $\hat{p} \leq \alpha_{\varepsilon}$ when in reality, $p>\alpha_{\varepsilon}$. The probability of such a false negative is given by

    \begin{equation*}
    \begin{aligned}
         \xi_p(N, \alpha_{\varepsilon}) & \equiv \mathbb{P}(\hat{p}\leq \alpha_{\varepsilon}) = \mathbb{P}(X\geq N(1-\alpha_{\varepsilon})) \\
        & = \sum_{i=\lceil N(1-\alpha_{\varepsilon})\rceil}^{N} \binom{N}{i}(1-p)^ip^{N-i}\ .
    \end{aligned}
    \end{equation*}

Our main result is stated in the following proposition:

\bigskip

\begin{proposition}
\label{theorem}
    For any $\gamma \in (0,1)$, the number of circuit repetitions necessary to $(1-\gamma)$-correctly identify that two points are not $\varepsilon$-neighbours is at least $\mathcal{O}\left(\frac{n^3}{\ln n}\right)$. In particular, this implies a total number of CSWAP gates of at least $\mathcal{O}\left(\frac{n^4}{\ln n}\right)$.
\end{proposition}

\bigskip

To prove this proposition, we will be using the following lemma and corollary. First let us define
    \begin{equation}
    \label{eq:ngamma}
        N(\gamma) = \frac{\ln(1/\gamma)}{(1-\alpha_{\varepsilon})\ln\left(\frac{1-\alpha_{\varepsilon}}{1-p}\right) + \alpha_{\varepsilon}\ln\left(\frac{\alpha_{\varepsilon}}{p}\right)}\ .
    \end{equation}  

\bigskip

\begin{lemma}
\label{lemma}
    There exists a $\tilde{\gamma}\in(0,1)$ as well as tuples $(\tilde{\alpha}_{\varepsilon},\tilde{p})$ such that for $\Tilde{N}\equiv N(\Tilde{\gamma})$, 

    \begin{equation*}
        \xi_{\tilde{p}}(\Tilde{N}, \tilde{\alpha}_{\varepsilon}) = \tilde{\gamma}\ .
    \end{equation*}
\end{lemma}

\medskip
\textit{Proof}: The Chernoff-Hoeffding inequality \cite[Lemma 4.7.2]{InformationTheory} states that for $0<q<\lambda<1$, $\mathbb{P}(X\geq N\lambda)\leq\exp\{-N\text{KL}(\lambda||q)\}$ where $\text{KL}(\lambda||q)=\lambda\ln\left(\frac{\lambda}{q}\right) + (1-\lambda)\ln\left(\frac{1-\lambda}{1-q}\right)$ is the Kullback-Leibler divergence between two Bernoulli variables with parameters $\lambda$ and $q$. Replacing $\lambda$ with $1-\alpha_{\varepsilon}$ and noting that $q=1-p$, we find that for $0<\alpha_{\varepsilon}<p<1$, 

\begin{equation}
\label{eq:uppercherhoeff}
    \xi_p(N, \alpha_{\varepsilon}) \leq \exp\{-N\text{KL}(\alpha_{\varepsilon}||p)\}\ .
\end{equation}
Thus, for any $\gamma\in(0,1)$ we have $\xi_p(N,\alpha_{\varepsilon})\leq\gamma$ for any $N\geq N(\gamma)$. Note that the KL divergence is non-negative and for our case non-zero (as $\alpha_{\varepsilon} < p$), so $N(\gamma)$ is well defined. 

We have the lower bound \cite[Lemma 4.7.1 and 4.7.2]{InformationTheory}
\begin{equation}
\label{eq:lowercherhoeff}
    \xi_p(N, \alpha_{\varepsilon}) \geq \frac{1}{\sqrt{2N}}\exp\{-N\text{KL}(\alpha_{\varepsilon}||p)\}
\end{equation}
for any $N\in\mathbb{N}^*, \alpha_{\varepsilon}, p$. Plugging in the expression for $N(\gamma)$ from (\ref{eq:ngamma}), we find that
\begin{equation}
\begin{aligned}
\label{eq:lbN}
    \xi_p(N(\gamma),\alpha_{\varepsilon}) & \geq \frac{1}{\sqrt{2\frac{\ln(1/\gamma)}{\text{KL}(\alpha_{\varepsilon}||p)}}}\exp\left\{\frac{\ln(\gamma)}{\text{KL}(\alpha_{\varepsilon}||p)} \text{KL}(\alpha_{\varepsilon}||p)\right\} \\
    & = \frac{\gamma}{\sqrt{2\frac{\ln(1/\gamma)}{KL(\alpha_{\varepsilon}||p)}}}\ .
\end{aligned}
\end{equation}
Setting the rightmost term in equation (\ref{eq:lbN}) equal to $\gamma$ will give us conditions on the values of $\text{KL}(\alpha_{\varepsilon}||p)$ for which the bound is sharp. We find that for 

\begin{equation*}
    \text{KL}(\alpha_{\varepsilon}||p) = 2\ln(1/\gamma)
\end{equation*}
$\xi_p(N(\gamma),\alpha_{\varepsilon}) = \gamma$. Note that for $\alpha_{\varepsilon}$ small and $p$ large, the KL-divergence explodes, so that for a large enough difference between $\alpha_{\varepsilon}$ and $p$, the equality holds for any $\gamma$ which concludes the proof. If we restrict the analysis to more common cases where $p$ is reasonably \footnote{By "reasonable" we mean values of $\alpha_{\varepsilon}$ and $p$ that are likely to occur in real experiments.} far from $1$, we find that for example for the tuple $(\tilde{\alpha}_{\varepsilon}=0.5, \tilde{p}=0.9)$, the KL-divergence is $0.51$, meaning that for $\tilde{\gamma} \approx 0.78$, and $(\tilde{\alpha}_{\varepsilon}, \tilde{p})$ as above, $\xi_{\tilde{p}}(N(\tilde{\gamma}),\tilde{\alpha}_{\varepsilon}) = \tilde{\gamma}$. 

\bigskip

\begin{corollary}
\label{corollary}
    There exist $\tilde{\gamma}\in(0,1)$ as well as tuples $(\tilde{\alpha}_{\varepsilon}, \tilde{p})$ such that for any $N \leq N(\Tilde{\gamma})\equiv\Tilde{N}$ and for any $\gamma\in (0,1)$ we can find reasonable $(\alpha_{\varepsilon},p)$ such that $\xi_{p}(N, \alpha_{\varepsilon}) \equiv \mathbb{P}(X\geq N(1-\alpha_{\varepsilon})) \geq \gamma$.
\end{corollary}

\medskip

In other words, there exists a minimal number $\tilde{N}$ of circuit repetitions such that for any precision parameter $\gamma$, there are couples $(\alpha_{\varepsilon}, p)$ such that the probability of identifying a false negative is greater than $\gamma$ for any number of circuit repetitions less than $\tilde{N}$.

\medskip

\textit{Proof}: This follows directly from Lemma \ref{lemma} by noting that for all $\gamma\geq\tilde{\gamma}$ as found in Lemma \ref{lemma}, reasonable tuples $(\alpha_{\varepsilon}, p)$ exist that fulfill the equality $\xi_p(N(\gamma), \alpha_{\varepsilon}) = \gamma$, and for $\gamma\leq\tilde{\gamma}$, the number of repetitions to attain $\gamma$ error must be at least $\Tilde{N}$.

\bigskip

\textit{Proof of Proposition \ref{theorem}}: From Corollary \ref{corollary} we know that for any $\gamma \in (0,1)$, the number of circuit repetitions necessary to $(1-\gamma)$-correctly identify that two points are not $\varepsilon$-neighbours is at least $ \tilde{N} = N(\tilde{\gamma})$. Note that at the end of circuit $\mathcal{U}_n$ the measurement outcome of the $d_n$ ancillary states is going to be one of the basis states of the ancilla register. Each basis state is associated with one pair of the input states, and measuring $0$ in the top most ancillary qubit is equivalent to measuring $0$ in the classic SWAP test. Adopting the notation from \cite{multiswap}, we denote by $p_{0ij}$ the probability of measuring the ancillary state $\ket{0ij}$ as in (\ref{eq:p0ij}). Note that $p_{0ij}\in[2^3/n^3, 2^4/n^3]$, $n\in \{2^k, k\geq 2\}$. We can then model one circuit run as a Bernoulli variable which returns the measurement outcome associated with the couple $(\ket{\phi_i}, \ket{\phi_j})$ with probability $p_{0ij}$, and any other outcome with probability $1-p_{0ij}$. We can thus replace $p$ from the previous sections with $p_{0ij}$.

Write 
\begin{equation*}
    \delta \equiv \mathds{1}_{|\phi-\psi|\leq\varepsilon}=\mathds{1}_{p_{0ij}\geq \frac{[(1-\varepsilon^2/2)^2+1]2^3}{n^3}}
\end{equation*}
 where $\varepsilon\in\mathbb{R}_+$. Then, 
 \begin{equation*}
    \alpha_{\varepsilon} = \frac{[(1-\varepsilon^2/2)^2+1]2^3}{n^3}
 \end{equation*}
 and for a false negative to be less than $\gamma$-probable, we need

\begin{equation*}
    N \geq \Tilde{N} = \mathcal{O}\left(\frac{n^3\ln(1/\tilde{\gamma})}{\ln n}\right)
\end{equation*}
circuit repetitions to $(1-\gamma)$ correctly identify that states $(\ket{\phi_i}, \ket{\phi_j})$ are not $\varepsilon$-neighbours. Since this is a lower bound for only one particular couple $(\ket{\phi_i}, \ket{\phi_j})$, and each circuit run only returns a measurement associated with exactly one couple, the total number of circuit runs to $(1-\gamma)$-correctly identify all non-$\varepsilon$-neighbours is at least $\mathcal{O}\left(\frac{n^3}{\ln n}\right)$.

\bigskip

\textit{Remark 1}: Note that by combining the lower and upper bound from (\ref{eq:lowercherhoeff}) and (\ref{eq:uppercherhoeff}), it becomes evident that the upper bound in (\ref{eq:uppercherhoeff}) is sharp asymptotically as $N\rightarrow \infty$.

\medskip

\textit{Remark 2}: From the above calculations, we see that the query complexity \textit{for all possible pairs} using the standard SWAP test is at least $\mathcal{O}(n^2\ln(1/\tilde{\gamma}))$. Since the naive multi-state SWAP test proposed in section \ref{sec:naivemultiswap} consists of running $\mathcal{O}(n^2)$ independent circuits (one for each possible distance), this implies a total number of CSWAP gates of $\mathcal{O}(n^4)$. The number of ancillary qubits is $\mathcal{O}(1)$, as opposed to $\mathcal{O}(\log n)$ for circuit $\mathcal{U}_n$.

\medskip

\textit{Remark 3}: The results of section \ref{sec:oraclecomplexity} are not directly comparable to the classical methods mentioned in section \ref{sec:ckassicaldistance}. While the brute-force algorithm exactly calculates all distances, and the kd-tree algorithm exactly calculates all relevant distances, the quantum method analysed in section \ref{sec:oraclecomplexity} is only concerned with approximately identifying $\varepsilon$-neighbours.

\section{Discussion}
\label{sec:discussion}
We have analysed the query and gate complexity of the multi-state SWAP test proposed in \cite{multiswap} for the purposes of creating an $\varepsilon$-graph, which are $\mathcal{O}\left(\frac{n^3}{\ln n}\right)$ and $\mathcal{O}\left(\frac{n^4}{\ln n}\right)$ respectively. Comparing these complexities directly to the time complexities of more common classical algorithms such as brute-force and kd-tree does not immediately yield a quantum advantage. This result makes the particular circuit $\mathcal{U}_n$ less suitable than other classical algorithms for classical data for the construction of an $\varepsilon$-graph, however the circuit might still be adapted for quantum data. 

A possible research avenue would be the search for better performing quantum distance algorithms. 
Although the algorithm from \cite{multiswap} might not be adapted for $\varepsilon$-graph construction, it could still be useful for other applications, in particular if the application calls for the preparation of all possible quantum state pairs in superposition in only two registers.

\nocite{*}
\bibliography{generic}
\end{document}